\documentclass[aps,prl,twocolumn,longbibliography]{revtex4-1}
\usepackage[utf8]{inputenc}
\usepackage{amsmath}
\usepackage{amsfonts}
\usepackage{amssymb}
\usepackage{graphicx}
\usepackage{float}
\usepackage[dvipsnames]{xcolor}
\usepackage{soul}
\usepackage{lipsum}

\begin{document}

\title{
Alignment induced re-configurable walls for patterning and assembly of liquid crystal skyrmions\\
} 

\author{
Ayhan Duzgun, Avadh Saxena 
} 
\affiliation{
Theoretical Division and Center for Nonlinear Studies,
Los Alamos National Laboratory, Los Alamos, New Mexico 87545, USA
} 
\author{
Jonathan V. Selinger
}
\affiliation{
Advanced Materials and Liquid Crystal Institute and Department of Physics,
Kent State University, Kent, OH,  44242, USA
}
\begin{abstract}

Skyrmions have attracted rapidly growing interest due to their topological properties and unique aspects for potential novel applications such as data storage and soft robotics. They can also serve as key elements for materials by design, self-assembly, and functional soft materials. While not real particles, these skyrmions behave like particles – they interact with each other and can be actuated by means of electric field, surface anchoring, and light. On the other hand, they are field configurations which have properties not possessed by real particles. Here, we show that, by means of alignment induced attractive and repulsive \emph{walls}, skyrmions in chiral nematic liquid crystals can be precisely controlled and programmed to serve as suitable building blocks for the realization of the above goals. Our work may stimulate new experimental efforts and concomitant applications in this direction.

\end{abstract}

\maketitle

{\it Introduction.} Richard Feynman pointed out that \emph{control of things on a small scale} would have great impact on customizing material properties and hence advance nanotechnology~\cite{feynman}. In this Letter, we propose that \emph{walls} produced by imposing alignment patterns can enable individual control of liquid crystal (LC) skyrmions, along the line of such purposes: Precise control of skyrmions in a LC cell can be achieved and employed to realize designer materials with desired properties at the micro-scale.

\begin{figure*}[t!]
\includegraphics[width=\linewidth]{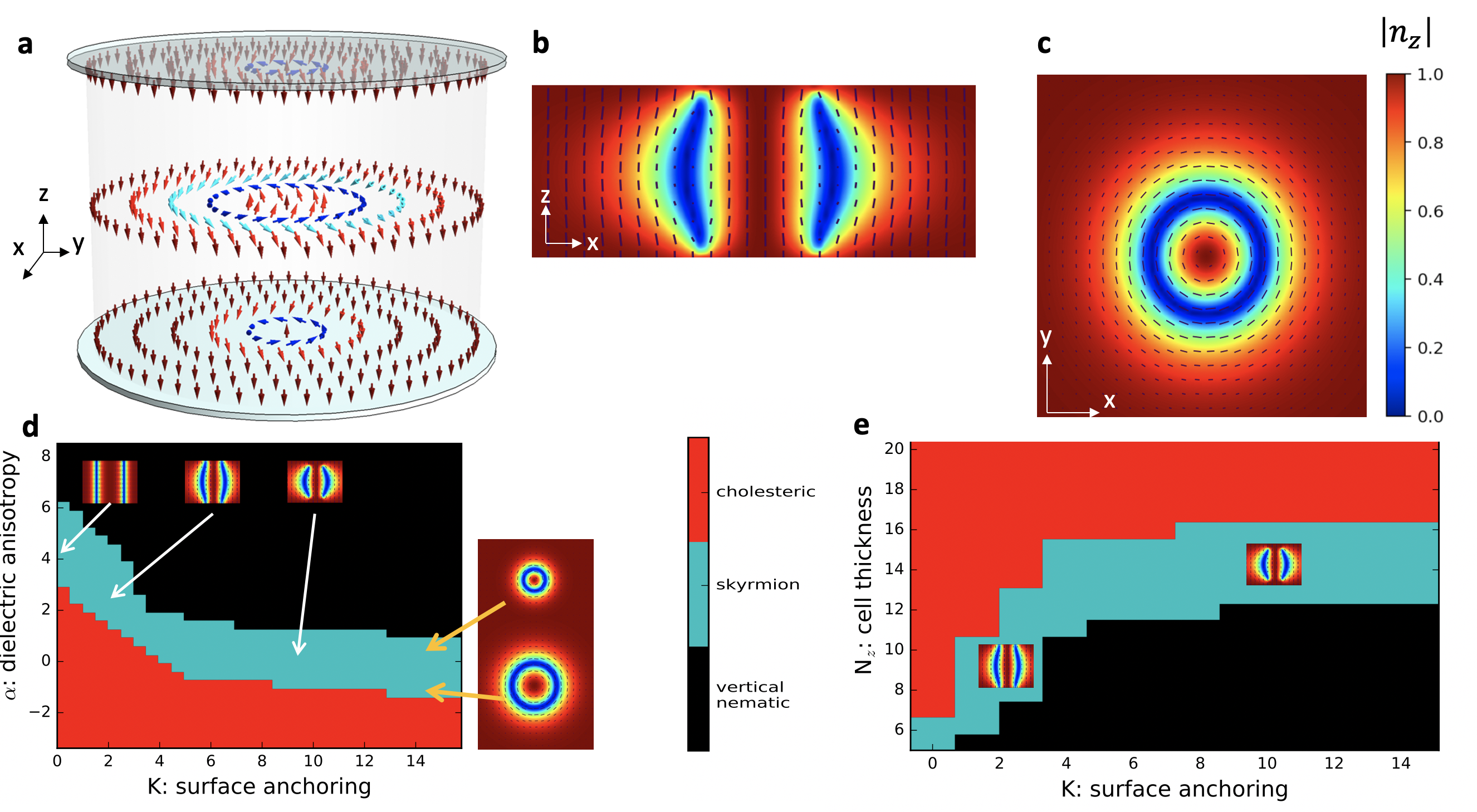}
\caption{Skyrmions realized in a chiral LC host confined between two parallel plates. (a) Vectorized director configuration of a skyrmion on three planes. (b) vertical and (c) horizontal cross sections of a skyrmion. Color mapping indicates the vertical component $|n_z|$ and black dashed lines indicate the projection of the director on the plane shown. (d) For the cell thickness $N_z=15$, field vs. surface anchoring phase diagram. (e) For no field ($\alpha=0$) cell thickness vs. surface anchoring phase diagram of skyrmions. (d,e) As the background alignment is varied, skyrmions exist between vertical nematic and cholesteric phases. The unit of $K$ is $K_0=LSq_0$ and the unit of $\alpha$ is $\alpha_0=LSq_0^2$. The unit of cell thickness is one lattice spacing equal to $1/40$ of the cholesteric pitch $p=2\pi/q_0$. }
\label{2D_3Dskyrmions}
\end{figure*}

Skyrmions are particle-like topological field configurations which were originally proposed as field configurations in particle physics~\cite{SKYRME1962556}, and later their realization in magnetic systems attracted much attention and extensive research has uncovered many complex features of magnetic skyrmions~\cite{Muhlbauer915,Nagaosa2013}. More recently, LC skyrmions have been realized as micron sized solitons in a chiral nematic material confined between two parallel substrates~\cite{Fukuda2011,leonov2014,bogdanov2003,Ackerman2014_2DSk,afghah2017,Guo2016}. See Fig.~\ref{2D_3Dskyrmions}~(a-c).

In LCs, full skyrmions are suitable for individual design and manipulation due to their particle-like properties; they exist as isolated objects~\cite{Ackerman2014_2DSk} and, when not tilted, they interact with other skyrmions like \emph{local} soft repulsive particles~\cite{Ackerman2015_assembly,Foster2019_SkBags} unlike other structures such as merons (half-skyrmions) that are accompanied by topological defects and are generally reported to exist as lattices~\cite{Hornreich_1978,Nych2017,active_merons,duzgun_SkPhase,Wang2018-surface-pattern}. Additional properties lacked by real particles provide an even richer set of tools as skyrmions can be generated and decimated at will, are flexible and readily deformable, and their interaction can be switched between isotropic-repulsive and directional-attractive~\cite{Ackerman2015_assembly}.

Novel mechanisms such as oscillating electric fields have been shown to drive particle-like excitations such as static skyrmions~\cite{Sohn6437} and solitary waves~\cite{electric-2,electric-1}. However, more localized control will better serve the purpose of design at the constituent level. In this work, we introduce a different approach enabling more precise and localized control and assembly of skyrmions into \emph{molecules}: instead of applying fields over a broad region, we generate regions with preferred alignment directions which act as repulsive or attractive walls. We impose the alignment of LC molecules by means of surface anchoring, electric field, and light. The reason we count light as a way to impose alignment is the following. Dimensionless  alignment strength can be expressed as the ratio of anchoring and field coupling coefficients to the natural twist $q_0=2\pi/p$, where $p$ is the helical pitch of the chiral nematic LC (see Eq.~\ref{FE} and the section \emph{Model}). Thus reducing $q_0$ increases the resulting alignment strength. For example, in the experiments of Sohn et al.~\cite{Sohn:19}, light is used to push skyrmions out of the exposed region as light chemically reduces the natural twist.   Our numerical simulations provide confirmation that modeling light exposure as a reduction in $q_0$ reproduces the observed behavior. More generally, we illustrate that any mechanism that impacts the helical pitch of the LC material can be used to generate attractive or repulsive regions.

We start by exploring the range of parameter space, as a reference point, in which we can realize skyrmions and then show how that range can be escaped to decimate skyrmions (a useful tool for dynamic manipulation). Next, we introduce repulsive and attractive walls produced by out of plane alignment and demonstrate how these soft and hard walls (obstacles) can be employed to guide and move skyrmions. Finally, we present simulations accompanied by an analytical calculation in the supplementary material (SM) to illustrate that obstacles with dynamically controllable size can be achieved when voltage across small electrodes is used as the wall generation mechanism. The size of these obstacles can be reduced all the way to zero by tuning the voltage, therefore this method allows even more precise and individual control over skyrmions.

Here, we emphasize that these mechanisms, due to their nature, can be best understood by viewing the videos included in the SM. This Letter investigates the case of \emph{upright} skyrmions stabilized by vertical alignment and lays the foundation for future work, in which we aim to extend our analyses to the case of topologically equivalent \emph{tilted} skyrmions~\cite{Ackerman2017_squirming}.

{\it Model.} Consider a \textit{LC cell} consisting of a chiral nematic LC confined between two parallel plates. We use Q-tensor representation to describe the average orientation of LC molecules at a certain position which is defined as $Q_{\alpha\beta}=\frac{3}{2}\left( \left<n_\alpha n_\beta\right> -\frac{1}{3} \delta_{\alpha\beta}  \right) $, where $n_\alpha$ are the components of the director field. In this notation Landau-de Gennes type free energy per unit volume, with single constant approximation, can be expressed as
\begin{align}
f&=\frac{1}{2}a \text{Tr}\left(Q^2\right)+\frac{1}{3}b \text{Tr}\left(Q^3\right)+\frac{1}{4}c\left[\text{Tr}\left(Q^2\right)\right]^2
\nonumber\\
&+\frac{1}{2}L\left(\partial _{\gamma }Q_{\alpha \beta }\right)\left(\partial _{\gamma }Q_{\alpha \beta }\right)- 2q_0L\epsilon_{\alpha \beta \gamma }Q_{\alpha \rho } \partial_{\gamma }Q_{\beta \rho }
\nonumber\\
&- \left( \Delta \epsilon E^2 + K\left[\delta(z)+\delta(z-N_z)\right]\right)Q_{zz}.
\label{FE}
\end{align}
The first line above is the thermal part. Below a critical temperature embedded in $a$, it yields a nematic phase with scalar order parameter $S$ determined by $a$, $b$, and $c$. The second line is the elastic part with coefficient $L$ and favors a natural twist $q_0$. Here $S$ is approximately constant everywhere except in defect regions.

Alignment is achieved via the third line which represents the homeotropic surface anchoring, at $z=0$ and $z=N_z$, and the external electric field $E$ applied in the $z$ direction, where $\Delta \epsilon$ is the dielectric anisotropy. Here $K$ is the coupling coefficient between $Q$ and a perfect alignment in the $z$ direction which enforces easy-axis alignment along $z$. As for the electric field, with the substitution of $\alpha=\Delta\epsilon E^2$, positive values of $\alpha$ indicate a material with easy-axis alignment while negative values indicate easy-plane alignment. The unit of $\alpha$ will be expressed in terms of $\alpha_0=LSq_0^2$ and the unit of $K$ in terms of $K_0=LSq_0$ so that they provide a comparison of  the cholesteric pitch $p=2\pi/q_0$ to the electric coherence length $\sqrt{LS/\alpha}$ and the extrapolation length $K/LS$, respectively. Larger values of $K/N_z$ and $\alpha$ represent greater {\it unwinding strength} against the natural twist $q_0$. Reduction of $q_0$ (light exposure) leads to greater $K/K_0$ and $\alpha/\alpha_0$ values and hence to a stronger alignment effect.

Numerical solution to the above free energy equation is obtained by simulations with over-damped relaxation dynamics, which is detailed in SM.

{\it Isolated skyrmions.} A strong easy-axis alignment yields a vertical nematic phase. With weaker alignment, a cholesteric phase is observed. An intermediate range of alignment strength allows isolated skyrmions with director profiles depicted in Fig.~\ref{2D_3Dskyrmions}~(a-c)~\cite{fingers2005,C9SM02033K,Guo2016,duzgun_SkPhase,afghah2017,tai2019surface}. The alignment strength is controlled by combinations of $K$, $N_z$, and $\alpha$. Fig.~\ref{2D_3Dskyrmions}~(d,e) show $K$-$\alpha$ phase diagram for $N_z=15$ and $K$-$N_z$ phase diagram for $\alpha=0$. To investigate the skyrmion behavior, we embed a generic skyrmion director profile inside a uniform background and let the system relax. If the alignment is too strong, the skyrmion shrinks and disappears. If the alignment is too weak then it expands and breaks into cholesteric stripes. The right panel of Fig.~\ref{2D_3Dskyrmions}~(d), shows how the skyrmion size is controlled as the background field is varied.

3D director profiles of the skyrmions vary depending on the alignment strength and mechanism. As seen in the insets of Fig.~\ref{2D_3Dskyrmions}~(d), if field alone is used a $z$-invariant structure forms. As $K$ is increased, the skyrmion takes a barrel shape (spherulite) and finally point or loop defects are formed. This behavior and further details have been discussed by Tai et al.~\cite{tai2019surface}.

The behavior illustrated here is key to generating and decimating skyrmions at will in any desired (programmed) fashion. Skyrmions \emph{written} by laser light~\cite{Ackerman2014_2DSk} can be decimated by locally escaping the suitable parameter range.

\begin{figure}[h]
\includegraphics[width=\linewidth]{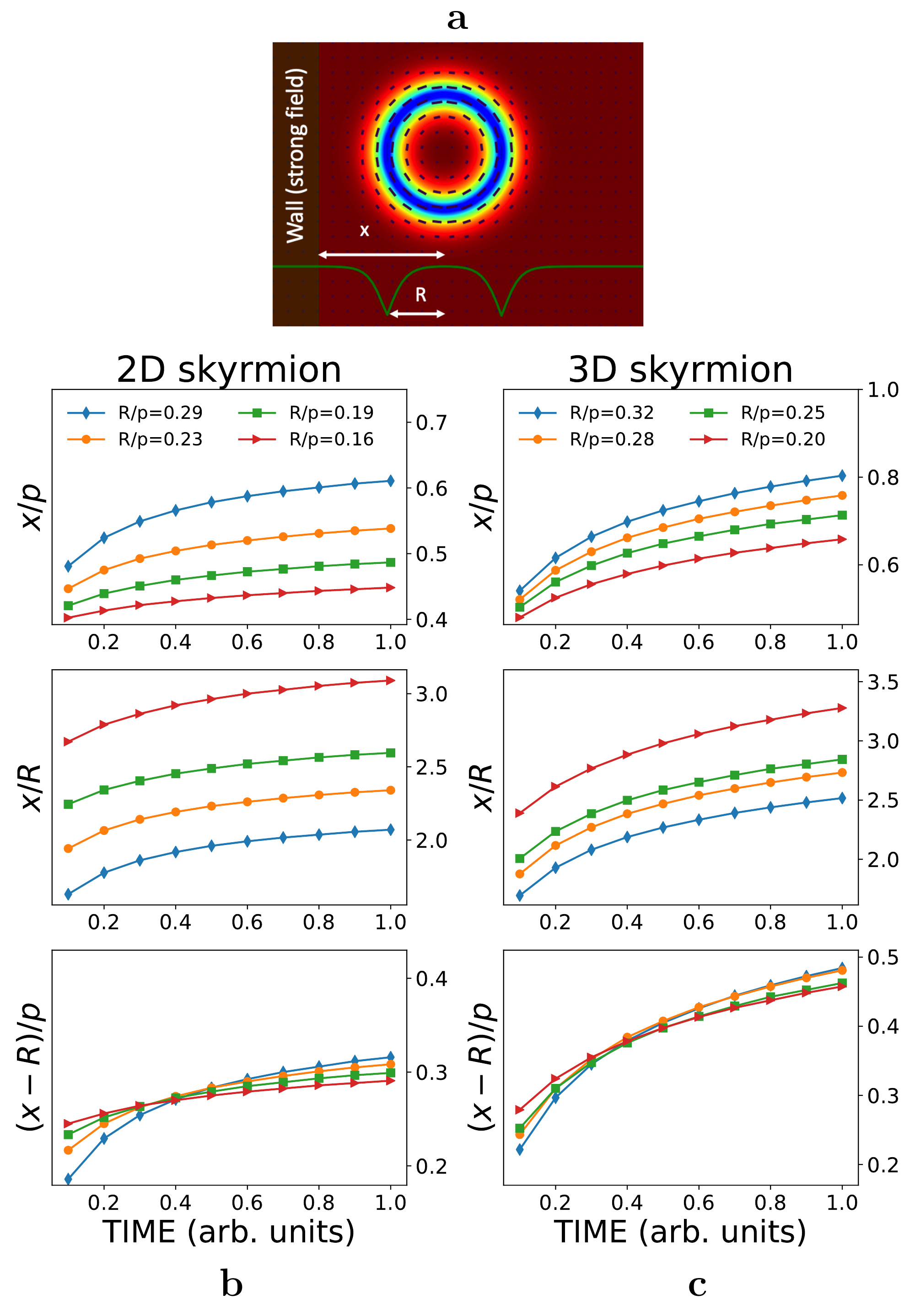}
\caption{ (a) A skyrmion near a wall produced by a strong vertical alignment. Green line indicates the absolute value of director $n_z$. Here $R$ is the skyrmion radius and $x$ is the distance from skyrmion center to the wall. Panels (b) and (c) plot the ratios $x/p$, $x/R$, and $(x-R)/p$ as a function of time for 2D and 3D skyrmions. 
}
\label{fig:wall-sk-sk}
\end{figure}

{\it Walls produced by strong vertical alignment.} The director profile of a skyrmion on a vertical cross section is basically a kink from its center to periphery in any radial direction as shown in Fig.~\ref{fig:wall-sk-sk}~(a) where the green line indicates the absolute value of $n_z$. For an isolated skyrmion embedded in a uniform far field, the $180^\circ$ rotation of the polar angle from the center to the periphery takes place over a characteristic length scale. That  length, determined by the natural twist coupled with the frustration in the system, is roughly a measure of how far away the skyrmion prefers to be from the perfectly vertical far field. Therefore, if we enforce a vertical alignment in a region we will essentially generate a wall that will push the skyrmions away. It is also clear that \emph{particle-like} repulsion of two skyrmions is due to the same reason$-$optimum relaxation length of the gradients of the polar angle to the minimum energy configuration. 

Now, we investigate the range of interaction between a skyrmion and a hard wall (produced by a very strong field) by placing a skyrmion near the wall and letting the system relax. In Fig.~\ref{fig:wall-sk-sk}~(b,c), we plot the ratios $x/p$, $x/R$, and $(x-R)/p$ vs. time for 2D and 3D skyrmions of various sizes where $x$ is the distance from the center of the skyrmion to the wall and $R$ is defined as the distance from the skyrmion center over which the first $90^\circ$ rotation of the polar angle occurs (radius of the blue circle). Thus the rotation from $90^\circ$ to $180^\circ$ occurs over $x-R$ because the hard wall maintains vertical alignment on the boundary. The size of the skyrmions is controlled by varying the alignment strength. In the case of a 2D skyrmion, we set $K=0$ and vary $\alpha$. In the case of a 3D skyrmion, we set $\alpha=0$ and vary $K$ (see SM for the parameters used.)

The results illustrate that the gradients of the polar angle are asymmetric around the horizontal direction. Furthermore, while the distance of rotation from $0^\circ$ to $90^\circ$ ($R$) varies greatly with the alignment strength, the rotation from $90^\circ$ to $180^\circ$ interestingly takes place over an almost constant distance $x-R$, which stays near $0.3p$ for 2D and $0.5 p$ for 3D skyrmions. Therefore, the range of interaction $x$ is not a certain multiple of $R$ but varies from two to several skyrmion sizes ($R$).

{\it Repulsive or Attractive Regions.} Confinement and controlled motion of skyrmions are essential features in order to custom design materials.  Surface anchoring is one way to impose boundary conditions for this purpose. Similarly, field alignment can produce regions with a desired director configuration. While surface anchoring requires pre-processing and is permanent, electric fields in principle can be turned on and off, moved, reoriented and varied with desired time dependence.

The wall mechanism described above can be used to generate repulsive regions. By the same token, attractive regions can be produced using alignment weaker than the background. The director profile of a skyrmion is approximately a vertical inner disk surrounded by a $\pi$-wall which in turn is surrounded by the vertical background. The non-vertical parts of a skyrmion will cost more energy when placed in a region with stronger vertical alignment due to larger restoring forces. Therefore, the non-vertical parts energetically prefer to overlap with weaker-alignment and the vertical parts prefer to overlap with stronger-alignment.

\begin{figure}[h]
\includegraphics[width=\columnwidth]{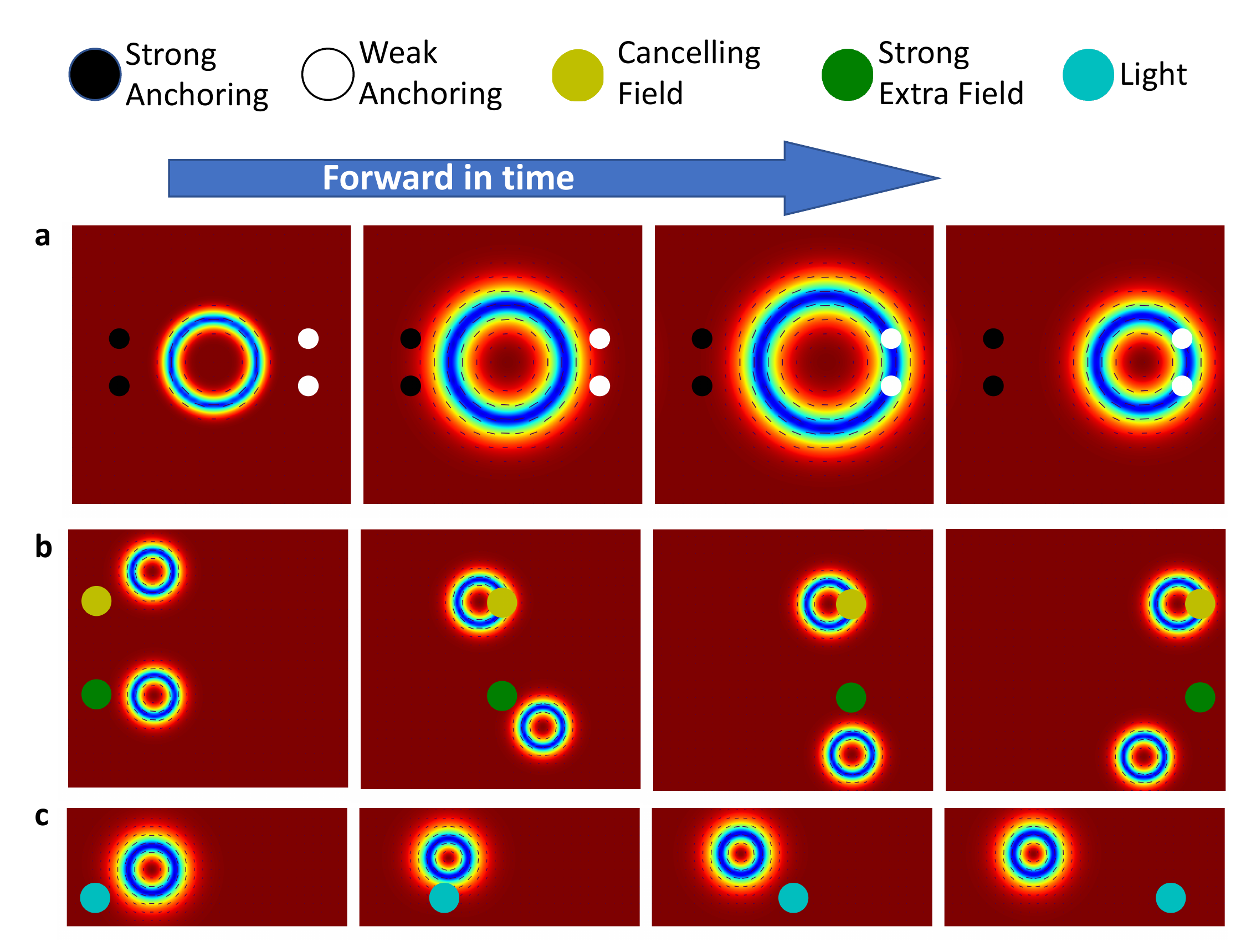}
\caption{ Snapshots from simulations showing skyrmions near moving or stationary spots made by modifications in surface anchoring, field, or light. (a) A skyrmion is swollen by lowering the background field, so it comes in contact with nearby spots. It is then repelled by strong anchoring sites (black) and attaches itself to the weak anchoring sites (white). (b) Skyrmions are dragged by a low field spot (yellow) and repelled by a high field spot (green). (c) Skyrmions are repelled by moving light. }
\label{fig:repulsive_attractive}
\end{figure}

Simulation snaphots in Fig.~\ref{fig:repulsive_attractive} are some arrangements which generate repulsive or attractive regions.  In (a), weaker and stronger surface anchoring regions are patterned inside a background anchoring strength of $K=2.6$ on the surfaces of a LC cell of thickness $15$. Black circles mark regions with strong surface anchoring, $K\approx 30$, and white circles mark regions with weak surface anchoring, $K\approx0.3$. Initially, with $\alpha=2.7$, the skyrmion is placed between these circles and then the background field is reduced so that the skyrmion swells and interacts with the black and white regions. Finally, the field is brought back to the initial value of $\alpha=2.7$ to deswell the skyrmion. In the final picture, the skyrmion ends up adhering to the weaker anchoring sites. In part (b), a LC cell of thickness $15$ with no surface anchoring is shown. The green spot denotes a region with stronger electric field of $\alpha=85$ in the $z$ direction and the yellow spot denotes a region in which the background field of $\alpha=5$ is canceled out by applying an electric field of equal magnitude in the opposite direction. Once these spots are moved towards the skyrmions, the green stronger-field spot repels and the yellow weaker-field spot attracts and drags the skyrmions. In part (c), we show snapshots demonstrating a different mechanism, light exposure. Exposure to light chemically increases the helical pitch of some LC materials~\cite{Sohn2018_cargo,Sohn:19}. To model the effect of light in simulations, we reduce the natural twist $q_0$ by 1.5 in the region of exposure. When we move that region towards a skyrmion, the skyrmion is repelled away from light. Our simulations confirm that reduction in $q_0$ is a valid mechanism to generate repulsive sites realized in the experiments mentioned above. All of these mechanisms can be used alone or in combination to generate patterned repulsive and/or attractive regions. (See SM for movies of each of the above mechanisms).

Can skyrmions be combined when squeezed together? As a test, we placed skyrmions between moving walls and reduced the volume between the walls [video in SM part (E)]. As the walls move closer like a \textit{trash compacter} no skyrmions fused. Interestingly, as the \textit{pressure} increases the skyrmions get smaller and smaller until they pop spontaneously.

{\it Fringe effect and very thin walls.} In the demonstrations above, we employed a uniform extra field produced in the form of a circular pillar as small as a fraction of a skyrmion which may seem experimentally difficult to achieve. In principle, the cross section of the pillar does not have to be small in order to push or drag skyrmions. However, in some cases fringe effects caused by the small electrodes generating the electric field across the LC cell will be important. Here we address the case of very small electrodes placed on both sides of the LC cell.

\begin{figure}[h]

\includegraphics[width=\columnwidth]{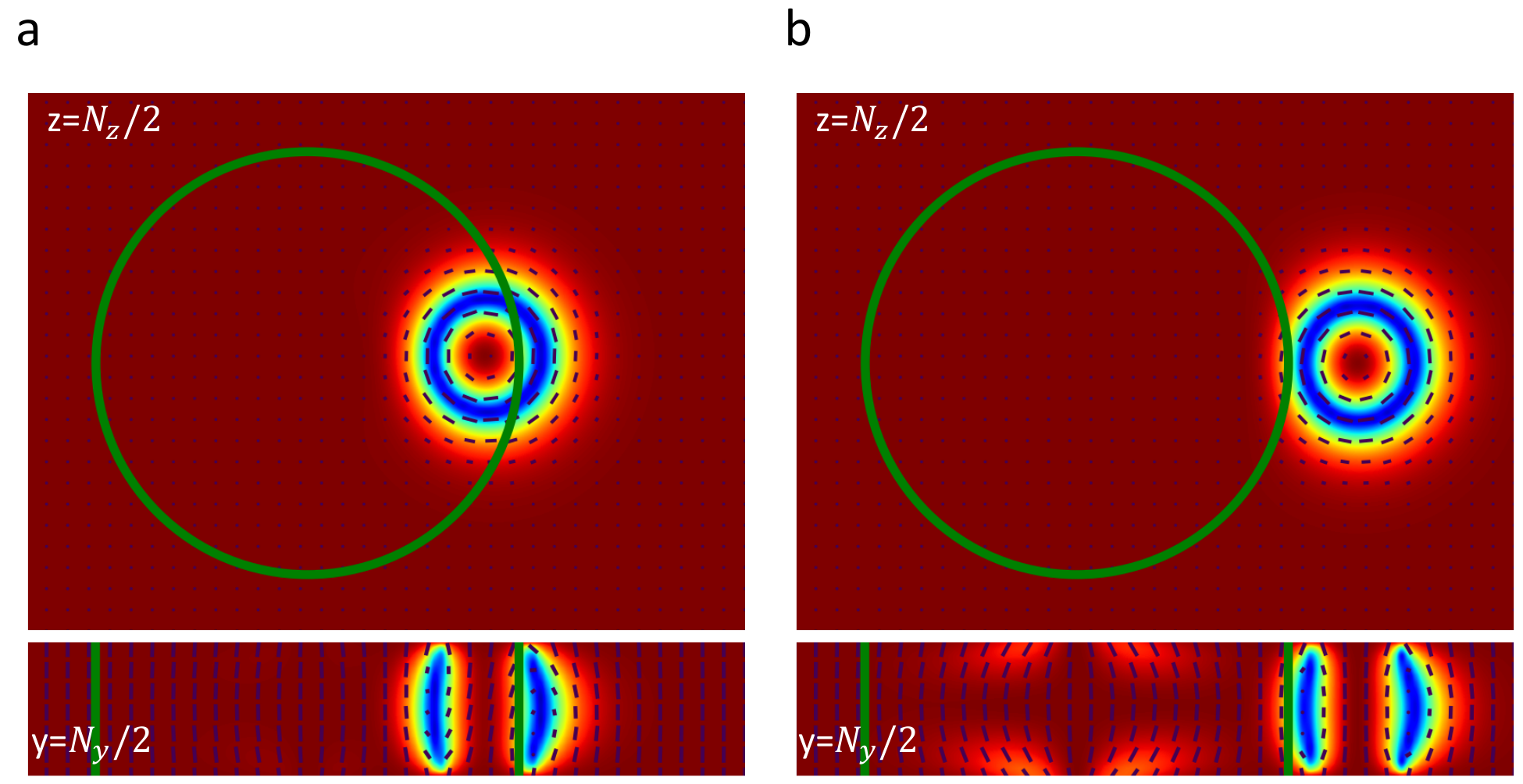}
\caption{{\bf Repulsive site generated by point electrodes:} Oppositely charged point charges are placed at positions $N_z/2\pm H$ above and below the mid-plane of the LC cell where $H=15$, $N_z=16$, $p=40$, $K/LSq_0=7.9$ and $\alpha/LSq_0^2=1.7$ with material parameters listed in SM. The green circle denotes the region with radius $R=2H=30$ outside which the field vanishes (see SM). When the skyrmions are driven by moving the wall region (green circle) an effective wall size is experienced depending on the potential difference (the greater the potential difference the greater the size). Top panels show horizontal cross section at $z=N_z/2$ and bottom panels show vertical cross section at $y=N_y/2$. The potential difference (charge) is chosen such that the extra field at the midpoint of the cell is equal to (a) 1 and (b) 6 times the background field.}

\label{fig:point_field}
\end{figure}

When a potential difference $\Delta V$ is applied between two such electrodes, charges of opposite sign and equal magnitude of $q=C\Delta V$ will  accumulate at the electrodes where $C$ is the capacitance. For very small size limit, we can safely assume that they are point charges. In Fig.~\ref{fig:point_field} we show how a field generated by such point charges can drive skyrmions. (The calculation of the field is shown in SM.) In part (a), a relatively weaker field is applied. In part (b), even when the extra field is much bigger than the background field leading to obvious deviation of the directors from the vertical direction, skyrmions are still successfully driven. Comparison of (a) and (b) panels of Fig.~\ref{fig:point_field} illustrates that the effective radius of the pillar depends on the potential difference applied across the LC cell. This  feature adds another useful \emph{knob} to the tool set to control skyrmions because changing the size of a repulsive (or attractive) pillar by means of varying the potential difference is much easier than changing the geometry and structure of the system and it allows dynamic control of the wall size. Furthermore, it helps overcome limitations on how small the walls can be made because we can reduce the electrode potential difference all the way to zero and achieve extremely small wall sizes.  

{\it Discussion.} We have demonstrated how electric field, surface anchoring and light can be used to generate repulsive and attractive regions in order to precisely design desired patterns and assemblies of skyrmions. The methods presented here can easily be modified or expanded to realize many diverse features for a variety of novel applications. In analogy to the letters or words in a book, skyrmions can be organized into artificial structures. As first steps, we have already numerically realized some useful and physics-rich applications including skyrmionic artificial spin ice~\cite{duzgun2019artificial} and dynamic switching of skyrmion lattice structures on substrates~\cite{duzgun2019commensurate}. We envisage that systems whose constituents are micron-sized skyrmions can be programmed to exhibit custom designed complex collective behaviors thanks to the possibility of individual creation and decimation of skyrmions at will and dynamic manipulation of obstacles, patterns, and even the way skyrmions interact with obstacles and other skyrmions. In addition to methods applied here to upright skyrmions, tilted skyrmions achieved by means of tilting or modulating background fields can add even richer features. Interesting skyrmion behaviors in the presence of field, light, and surface anchoring are continuously being discovered. Original experimental studies~\cite{electric-1,Sohn:19,electric-2,bukusoglu-annurev} that appeared recently also indicate that we are at the verge of significant progress in realizing functional materials whose constituents are skyrmions. We believe that the ideas presented here will inspire a new direction for the realization of designer skyrmion materials and motivate experimental and theoretical efforts aiming to investigate the underlying physics in more detail. Furthermore, similar methods and mechanisms can potentially be applied in magnetic systems as well~\cite{Lin2013, chiral_sk_in_gradients, Leonov_2016, leonov-spintronics, skyrmion-gates}.

\begin{acknowledgments}
This work was supported in part by the U.S. Department of Energy and in part by NSF Grant DMR-1409658.
\end{acknowledgments}

\onecolumngrid

\newpage

\newcommand{\pillarfigs}{.}
\begin{center} 
{\Large\bf Supplementary Material}
\vspace*{.2in}
\end{center}

\section{Electric field inside dielectric slab placed between point electrodes}

A mechanism to generate walls that repel \emph{upright} skyrmions stabilized in a chiral nematic liquid crystal (LC) cell is to produce extra vertical alignment by means of strong  additional  field. Here we explore how this can be done when the field cross section is very small and its magnitude and direction are not uniform. 

We investigate the alignment effect of applying a potential difference between two very thin electrodes placed on both sides of liquid crystal cell. The LC cell itself is filled with a chiral nematic material with positive dielectric anisotropy that hosts full skyrmions. This requires, apart from the potential we will apply, already existing vertical alignment produced by homeotropic surface anchoring or a background electric field perpendicular to the plane of the cell because skyrmions are embedded in a uniform vertical far-field director. To generate a hard wall (a region with strong vertical alignment) the field from the applied potential should be much bigger than these existing alignment factors. 

We will assume that the tips of the electrodes are very small spheres thus the charge accumulated at the tips will be modeled as point charges of equal magnitude but apposite sign symmetrically placed around the LC cell (Fig. 1). 

First let's consider only the potential from one electrode perform the calculation with very slight modification of the method introduced by~\cite{sometani}.  

\section*{Potential inside dielectric slab near a point charge}

A point charge $q$ is placed near a dielectric slab with dielectric constant $\epsilon$. If the dielectric medium was semi-infinite as in~\cite{jackson}, rather than a slab, the potential inside the dielectric medium would be equivalent to the potential due to the original charge plus one image charge overlapping with the original charge with magnitude $q'=-\beta q$ where $\beta=(\epsilon-\epsilon_0)/(\epsilon+\epsilon_0)$. However for the slab, another image charge is needed in order to satisfy the boundary conditions on the second plane. In turn additional charges will be required because boundary conditions on the first plane won't be satisfied anymore. Then the additional charges will require more image charges and so on.

\begin{figure}[h]
\center
\includegraphics[width=0.8\columnwidth]{\pillarfigs/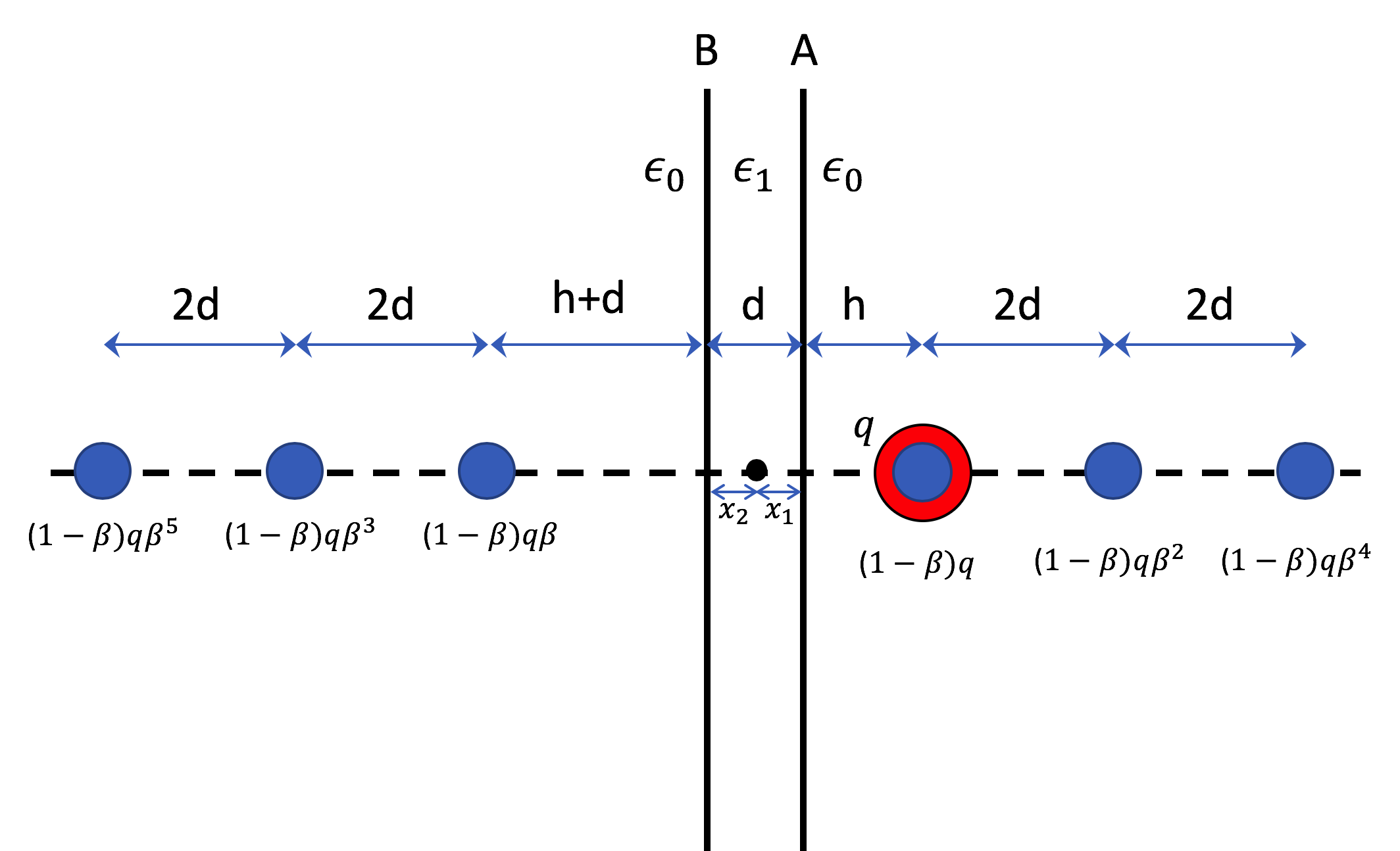}
\caption{{\bf Calculation of the potential inside a slab by  the method of image charges:} Red circle represents the original free charge $q$ placed at a distance $h$ from the right side of the slab of thickness $d$. Blue circles denote the infinite series of image charges at the positions shown. $\beta=(\epsilon-\epsilon_0)/(\epsilon+\epsilon_0)$. }
\label{fig1}
\end{figure}

Therefore it is seen that the infinite set of image charges shown in Fig.~\ref{fig1} is one way to get the solution for the potential at a point between the planes. The first image charge overlaps with the original charge. Thus to calculate the second image charge we can use the total charge $q-\beta q=(1-\beta) q$ as \emph{felt} at the location of the charge that is . At each step, the next image is found by locating it symmetrically about the plane of the corresponding surface charges and multiplying the magnitude of the charge by $\beta$. At a point on the axis of symmetry located at distances $x_1$ from the surface closer to the charge $q$ and $x_2$ from the surface away from $q$, the potential due to the charges on the right side is 

$$V_R = \frac{(1-\beta) q}{4\pi\epsilon_0 (x_1+h)} +  \frac{(1-\beta) q \beta^2}{4\pi\epsilon_0 (x_1+h+2d)} + ...  = \frac{(1-\beta) q}{4\pi\epsilon_0} \sum_{n=0}^\infty \frac{\beta^{2n}}{x_1+h+2nd} .$$
Similarly, for the charges on the left hand side 
$$V_L = \frac{(1-\beta) q \beta}{4\pi\epsilon_0 (x_2+h+d)} +  \frac{(1-\beta) q \beta^3}{4\pi\epsilon_0 (x_2+h+3d)} + ...  = \frac{(1-\beta) q}{4\pi\epsilon_0} \sum_{n=0}^\infty \frac{\beta^{2n+1}}{x_2+h+(2n+1)d} .$$
The potential due to all charges are then
\begin{align}
 V= \frac{(1-\beta) q}{4\pi\epsilon_0} \sum_{n=0}^\infty  \left(  \frac{\beta^{2n}}{x_1+h+2nd} +  \frac{\beta^{2n+1}}{x_2+h+(2n+1)d}   \right) 
\end{align}
At a radial distance $r$ from the central axis of the slab (line connecting the charges) is
\begin{align}
V&= \frac{(1-\beta) q}{4\pi\epsilon_0} \sum_{n=0}^\infty  \left(  \frac{\beta^{2n}}{  \sqrt{r^2 + (x_1+h+2nd)^2} } +  \frac{\beta^{2n+1}}{\sqrt{ r^2 + (x_2+h+(2n+1)d)^2} }   \right)
\end{align}

\section*{Potential from two point charges located on sides of a LC cell}

Using the result of the single charge case, potential inside the dielectric slab from two oppositely charged electrodes can be written, after swapping $x_1$ and $x_2$ for $-q$, as 

\begin{eqnarray}
V= \frac{(1-\beta) q}{4\pi\epsilon_0} \sum_{n=0}^\infty  \bigg (  \frac{\beta^{2n}}{  \sqrt{r^2 + (x_1+h+2nd)^2} } +  \frac{\beta^{2n+1}}{\sqrt{ r^2 + (x_2+h+(2n+1)d)^2} } \nonumber \\
-\frac{\beta^{2n}}{  \sqrt{r^2 + (x_2+h+2nd)^2} } -  \frac{\beta^{2n+1}}{\sqrt{ r^2 + (x_1+h+(2n+1)d)^2} }
\bigg)
\end{eqnarray}
which then can be simplified by combining the terms with denominators containing $x_1$ or $x_2$:
\begin{eqnarray}
V= \frac{(1-\beta) q}{4\pi\epsilon_0} \sum_{n=0}^\infty  \bigg(  \frac{(-\beta)^{n}}{  \sqrt{r^2 + (x_1+h+nd)^2} } -  \frac{(-\beta)^{n}}{  \sqrt{r^2 + (x_2+h+nd)^2} }
\bigg)
\end{eqnarray}
where $x_1$ is the distance to the plane closer to $+q$ and $x_2$ is the distance to the plane closer to $-q$.

\section*{LC cell between electrodes}

A similar result to the above solution is obtained by~\cite{deng-ke} for a different arrangement of LC cell and coating. We argue(expect?) that typical arrangements of LC material, glass substrates, coatings etc. will yield similar potentials as long as the system consists of planar surfaces. Therefore as a generic example we use the dielectric slab described above.

We position the slab of thickness $d$ at the origin. Two point charges $\pm q$ are placed at $z=\mp H$ respectively. Distances from the point where the field calculated are $x_1+h=H-z$ and $x_2+h=H+z$. Then $E_z=-\partial_z V$ and $E_r=-\partial_r V$ reads
\begin{equation}
E_z=\frac{(1-\beta) q}{4\pi\epsilon_0} \sum_{n=0}^\infty \left( \frac{(-\beta )^n (nd+H-z)}{\left[(nd+H-z)^2+r^2\right]^{3/2}}+\frac{(-\beta )^n (nd+H+z)}{\left[(nd+H+z)^2+r^2\right]^{3/2}} \right)
\label{eq:Ez}
\end{equation}
\begin{equation}
E_r=\frac{(1-\beta) q}{4\pi\epsilon_0} \sum_{n=0}^\infty \left( -\frac{(-\beta )^n r }{\left[(nd+H-z)^2+r^2\right]^{3/2}}+\frac{(-\beta )^n r}{\left[(nd+H+z)^2+r^2\right]^{3/2}} \right)
\label{eq:Er}
\end{equation}

\begin{figure}
\includegraphics[width=\columnwidth]{\pillarfigs/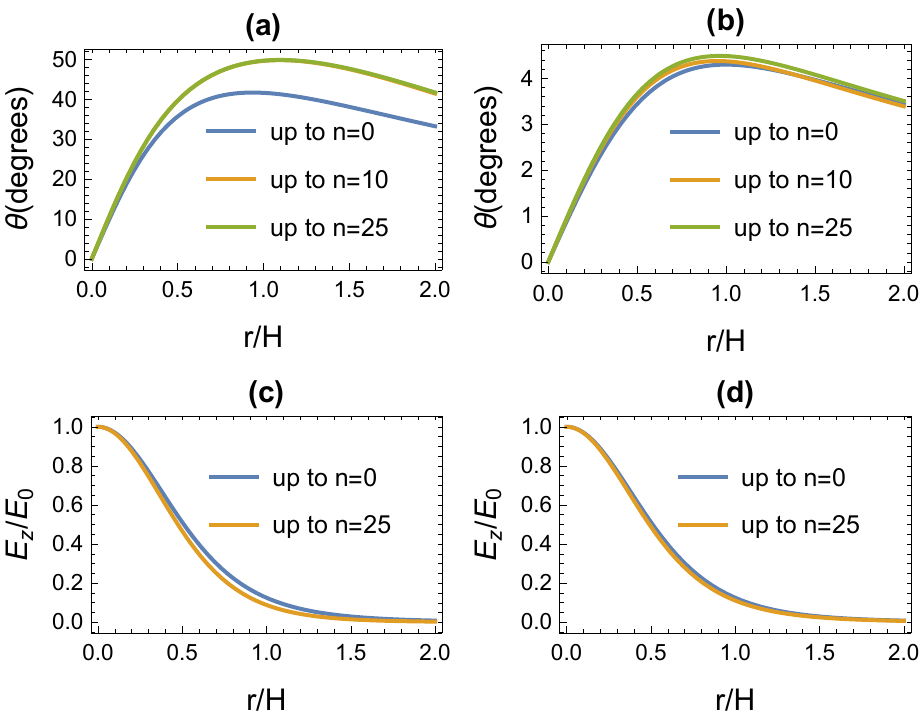}
\caption{ {\bf Comparison of fields calculated from Eqs.~\ref{eq:Ez} and~\ref{eq:Er} using first $n$ terms. }Top: Angle of electric field measured from $z$-axis for (a) $d=H$. (b) $d=0.1H$. Bottom: $E_z(z=0)$ normalized by $E_0=E_z(z=0,r=0)$ up to indicated number of terms for (c) $d=H$. (d) $d=0.1H$.  }
\label{fig2}
\end{figure}

Next we compare the full sum of series to only the first term substituting dielectric coefficient $\epsilon =10$(anisotropy ignored for simplicity) and $\beta=(10-1)/(10+1)\approx 0.8$. In Fig.~\ref{fig2} we plot $E_z$ and the deviation of field lines from the vertical direction i.e $\theta=\arctan (E_r/E_z) $ as a function of $r/H$ for two different values of the cell thickness $d$.

\subsection*{Conclusion}
Regardless of how many terms we use to calculate the fields a more important feature is that $E_z$ decays to zero within a radius $r\leq 2H$. This enables to create pillars made of electric field with size less than $2H$. The actual effective radius will depend on various parameters and it can be directly controlled by varying the potential difference across the electrodes which enables us to achieve pillar sizes that can be varied from very small to big values.

As for how much approximation is appropriate, Fig~\ref{fig2} illustrates that, the first terms in  Eq.~\ref{eq:Er}  and~\ref{eq:Ez} capture the essential features of the exact sum. In the simulations, since we normalize the extra field by the value at the origin (middle of the dielectric slab), the first term $E_z$ is very close to the infinite sum (Fig.~\ref{fig2}-(c),(d)). Therefore we can say that fields from two point charges will be sufficient to capture the physics, although there is not much difficulty in including more terms for numerical calculations.

\bibliography{MovingSk2.bib}

\end{document}